\documentclass[12pt, twoside, leqno]{article}

\usepackage{lmodern}
\usepackage{amsmath,amsthm}
\usepackage{amsfonts}
\usepackage{amssymb,latexsym}
\usepackage{enumerate}
\usepackage[T1]{fontenc}
\usepackage{geometry}
\usepackage{marginnote}
\usepackage{graphicx}
\usepackage{multirow}
\usepackage{caption}
\usepackage{subfig}

\DeclareMathOperator{\lehmer}{lc}

\theoremstyle{definition}
\newtheorem{thm}{Theorem}[section]

\newtheorem{defin}[thm]{Definition}

\numberwithin{equation}{section}

\begin{document}

\baselineskip=17pt

\title{Generic Networks of Votings}

\author{Ewa Zawi\'slak-Sprysak \\
USUMA GmbH Institut f\"ur Marktforschung und Sozialforschung \\
E-mail: ewa.sprysak@usuma.com \\
\\
Pawe\l{} Zawi\'slak\\
Department of Mathematics and Mathematical Economics \\
SGH Warsaw School of Economics \\
E-mail: pzawis@sgh.waw.pl \\
}

\date{\today}

\maketitle

\renewcommand{\thefootnote}{}
\footnote{\emph{JEL Classification System}: C02, C15, D71, D85.}



\renewcommand{\thefootnote}{\arabic{footnote}}
\setcounter{footnote}{0}

\begin{abstract}
In this paper, we analyse results of the $15^{\textrm{th}}$ International Henryk Wieniawski Violin Competition by comparing the properties of its results network to the properties of \emph{generic networks of votings}.

Suppose that a competition \textbf{Comp} is given. In this competition, $k$ contestants are rated by $n$ jurors. Suppose that the Borda count is used as a voting method, i.e. every juror gives $k$ points for the best contestant, $k-1$ points for the second best contestant, and so on. In particular, the worts contestant gets $1$ point. 

For such a competition, we create a weighted network $N(\textrm{\textbf{Comp}})$ in the following way. The node set of $N(\textrm{\textbf{Comp}})$ corresponds to jurors and the link set of $N(\textrm{\textbf{Comp}})$ consists of all links 
$\{J_{s},J_{t}:s\neq t\}$. For link $l_{st}$ connecting nodes $J_{s}$ and $J_{t}$, we assign the weight $w(l_{st})=w_{st}$, where 
\begin{displaymath}
w_{st}=\mathcal{LF}_{2}(\alpha_{s}\alpha_{t}^{-1}) \textrm{.}
\end{displaymath}
Here $\mathcal{LF}_{2}$ is a Lehmer norm on the permutation group $S_{k}$, whereas $\alpha_{s}$ and $\alpha_{t}$ denotes the votes of jurors $J_{s}$ and $J_{t}$, respectively. In particular, for $i=1,2,\ldots,k$, $\alpha_{s}(i)$ is the number of points given to the $i$-th contestant by juror $J_{s}$. The similar holds for juror $J_{t}$. Note that $\alpha_{s}$ and $\alpha_{t}$ can be considered as elements of $S_{k}$.

Suppose now that the probability measure $\mathbb{P}$ is given on space $V$ of all possible votings of a single juror, i.e. on space $S_{k}$. Suppose that every juror votes independently according to $\mathbb{P}$. We repeat such a voting process $100$ times and for every $j=1,2,\ldots,100$, we create a network $N_{j}$ in the way described above. 

In this paper, we compare some statistical properties of networks $N_{j}$, for probability measures $\mathbb{P}$ being the convex combinations of two Dirac probability measures and a uniform probability measure, to the properties of network of jurors' votings in the $2016$ Wieniawski Competition.
\end{abstract}

\section{Introduction}\label{sec:Introduction}

There where many controversies concerning the results of the $15^{\textrm{th}}$ International Henryk Wieniawski Violin Competition (2016). Both Gazeta Wyborcza, one of the most popular Polish newspaper, see \cite{D}, and Ruch Muzyczny, the most significant Polish music journal, see \cite{JC}, raised the possibility that the jurors of the most recent Wieniawski Competition formed cliques.

The results of the 2016 Wieniawski Competition were analysed in \cite{SZ}, where they were compared to the results of $16^{\textrm{th}}$ (2010) and $17^{\textrm{th}}$ (2015) International Chopin Piano Competitions. The metods of network theory, see for example \cite{J}, \cite{Ne}, \cite{NBW} and \cite{W}, were applied to compare the voting results of the three aforementioned music competitions. For these three competitions weighted networks, see \cite{H}, $W^{2016}$, $C^{2010}$ and $C^{2015}$ were created and some numerical properties of these network were compared.    

Weighted networks are used in biology, see \cite{H}, in stock markets analysis, see \cite{CGQS} and \cite{GDKO}, as well as in the studies on the structural and functional organisation of the human brain, see \cite{PF}. Usually, the weight of link $l_{st}$ connecting nodes $s$ and $t$ are given by some kind of a correlation coefficient related to some rankings or processes asocciated to nodes. In \cite{SZ}, the weights $w_{st}$ of links $l_{st}$ were given by $w_{st}=\tau_{st}$, where $\tau_{st}$ were \emph{Kendall’s $\tau$ coefficients}, 
see \cite{K1}, \cite{K2} and \cite{A}, of the voting results of jurors $J_{s}$ and $J_{t}$. For stock market networks, this correlation was measured by the \emph{Pearson correlation coefficient}, see \cite{BS}.

In the case of rankings, many measures of disarray has been studied in the literature. In the case of the \emph{Borda count} method of voting, see \cite{HM}, \cite{Nu} and \cite{O} for a description of this method, votes can be regarded as elements of permutations group $S_{k}$ ($k$ is a number of contestants). The best known measures of dissarray are \emph{Kendall's $\tau$ correlation coefficient} and \emph{Spearman's $\rho$ correlation coefficient}, see \cite{S}, as well as metric measures such as \emph{Kendall distance}, \emph{Spearman distance}, \emph{Hamming distance} and \emph{Footrule distance}, see \cite{DG} and \cite{QY}. The weighted versions of the Kendall distance and the Footrule distance were considered in \cite{KV} and \cite{PP}. 

The classical measure of dissarray mentioned above has such a property that changing the first two positions in the Borda ranking has the same impact on the measure as changing the last two positions. On the other hand, the weighted generalisations of the Kendall distance and the Footrule distance proposed in \cite{KV} fail to be metrics (for some choices of weights). 

In \cite{Z}, the \emph{Lehmer factorial norm} is considered. This is a symmetric, right--invariant norm on the permutation group $S_{n}$ satisfying the triangle inequality and thus determining the metric on $S_{n}$. Additionally, this norm allows for distinguishing changes in the first positions and in the last positions of rankings. We use this norm to create networks related to the results of the $15^{\textrm{th}}$ International Henryk Wieniawski Violin Competition. We compare the properties of these networks to the average properties of networks related to randomly choosen results of a competition with the same numbers of contestants and jurors. 

This article is organised in the following way. In Section \ref{sec:Notation}, we present basic definitions concernig permutation groups and we set the notations used in this paper. We also recall the definition and basic properties of the Lehmer factorial norm. In Section \ref{sec:Wieniawski}, we analyse the results of the $15^{\textrm{th}}$ International Henryk Wieniawski Violin Competition using the network approach. In Section \ref{sec:Generic}, we describe the procedure of generating random networks of votings. These \emph{generic networks} are used later in Section \ref{sec:Fitting} for determinig the model of generating random networks with properties best fitting the properties of the network related to the results of the $15^{\textrm{th}}$ International Henryk Wieniawski Violin Competition. Section \ref{sec:Conclusions} contains conclusions and open questions related to the subject of these studies. All tables and figures are included in Section \ref{sec:Tables}.

\section{Basic definitions and notations, the Lehmer norm}\label{sec:Notation}
In this section, we presrent some basic definitions used in this paper and we set some notations. We also refer to the definition and basic properties of the Lehmer factorial norm.

For a natural number $n>0$ by $[n]$, we denote the set $\{1,2,\ldots,n\}$ and by $S_{n}$ -- the group of all permutations of $[n]$.
Permutation $\sigma \in S_{n}$ is denoted by 
\begin{displaymath}
\sigma=(\sigma(1),\sigma(2),\ldots,\sigma(n)) \textrm{.}
\end{displaymath} 
In particular $\varepsilon_{n}=(1,2,\ldots,n)$ denotes the identity permutation. 

By $\sigma^{-1}$, we denote the inverse permutation to $\sigma$, and by $\sigma\tau$ -- the composition of $\sigma$ and $\tau$, defined by $(\sigma\tau)(i)=\sigma(\tau(i))$ for $i=1,2,\ldots,n$. By $\bar{\sigma}$, we denote the reverse permutation to $\sigma$ given by $\bar{\sigma}(i)=\sigma(n+1-i)$ for $i=1,2,\ldots,n$.

For $s=1,2,\ldots,n-1$ let 
\begin{displaymath}
\alpha_{n}^{s}=(1,2,\ldots,s-1,s+1,s,s+2,\ldots,n) \textrm{,}
\end{displaymath}
so $\alpha_{n}^{s}$ is adjacent transposition, $(s,s+1)$ in the cycle notation.

For permutation $\sigma \in S_{n}$, its \emph{Lehmer code} $\lehmer(\sigma)$, see \cite{L1}, \cite{L2} and \cite{G}, is defined by
\begin{displaymath}
\lehmer(\sigma)=[c_{1}(\sigma),c_{2}(\sigma),\ldots,c_{n}(\sigma)]
\end{displaymath}
where numbers $c_{i}(\sigma)$ are given by
\begin{displaymath}
c_{i}(\sigma)=\left|\{j \in [n]: j>i \textrm{ and } \sigma(j)<\sigma(i)\}\right|
\end{displaymath}
for $i=1,2,\ldots,n$.

\begin{defin}[Definition 3.4 in \cite{Z}]\label{def:LehmerNorm}
Let $\sigma \in S_{n}$ be a permutation with the Lehmer code 
\begin{displaymath}
\lehmer(\sigma)=[c_{1}(\sigma),c_{2}(\sigma),\ldots,c_{n}(\sigma)] \textrm{.}
\end{displaymath}
\emph{Lehmer factorial norm} $\mathcal{LF}_{2}:S_{n} \to \mathbb{N}$ is given by 
\begin{displaymath}
\mathcal{LF}_{2}(\sigma)=\sum_{i=1}^{n}\left[2^{n-i}-2^{n-i-c_{i}(\sigma)}\right] \textrm{.}
\end{displaymath}
\end{defin}

In the next theorem, we refer to some basic properties of the Lehmer norm.

\begin{thm}[Theorem 3.6 in \cite{Z}] \label{thm:LehmerNorm} 
Norm $\mathcal{LF}_{2}$ satisfies the following:
\begin{enumerate}[(i)]
\item\label{thm:LehmerNormMinimal} $\mathcal{LF}_{2}(\varepsilon_{n})=0$ is minimal and $\varepsilon_{n}$ is the only permutation with this property.
\item\label{thm:LehmerNormMaximal} $\mathcal{LF}_{2}(\bar{\varepsilon}_{n})=2^{n}-(n+1)$ is maximal and $\bar{\varepsilon}_{n}$ is the only permutation with this property.
\item\label{thm:LehmerNormOrder} $\mathcal{LF}_{2}(\alpha_{n}^{s})=2^{n-1-s}$ for $s=1,2,\ldots,n-1$, and therefore
\begin{displaymath}
\mathcal{LF}_{2}(\alpha_{n}^{1})>\mathcal{LF}_{2}(\alpha_{n}^{2})>\ldots>\mathcal{LF}_{2}(\alpha_{n}^{n-1}) \textrm{.}
\end{displaymath}
\item\label{thm:LehmerNormSymmetry} $\mathcal{LF}_{2}(\sigma)=\mathcal{LF}_{2}(\sigma^{-1})$ for all $\sigma \in S_{n}$.  
\item\label{thm:LehmerNormTriangle} $\mathcal{LF}_{2}(\sigma\tau) \leq \mathcal{LF}_{2}(\sigma)+\mathcal{LF}_{2}(\tau)$ for all $\sigma,\tau \in S_{n}$.  
\end{enumerate}
\end{thm}

Note that properties (\ref{thm:LehmerNormMinimal}), (\ref{thm:LehmerNormSymmetry}) and (\ref{thm:LehmerNormTriangle}) imply that 
$\mathcal{LF}_{2}$ determines the metric on $S_{n}$. Indeed, the function $d_{\textrm{L}}:S_{n}\times S_{n}\to\mathbb{N}$ given by 
\begin{displaymath}
d_{\textrm{L}}(\sigma,\tau)=\mathcal{LF}_{2}(\sigma\tau^{-1})
\end{displaymath}
is a metric. We call it the \emph{Lehmer distance}.

Note also that $d_{\textrm{L}}$, considered as a distance on rankings, distinguishes changes in high places in the competitions from changes in low places. 

\section{Results of the $15^{\textrm{th}}$ Wieniawski Competition. A network approach}\label{sec:Wieniawski}
In this section, we analyse the results of the $15^{\textrm{th}}$ International Henryk Wieniawski Violin Competition using the network approach.

In this paper we, consider simply undirected networks. A \emph{(simply undirected) network} is a pair $N=(N(N),L(N))$ consisting of set $N(N)$ of \emph{nodes}, usually finite, and set $L(N)$ of \emph{links}, where every link $l\in L(N)$ is a subset of $N(N)$ consisting of two different elements. Networks are often called \emph{graphs} in the literature, nodes and links - \emph{vertices} and \emph{edges}, \emph{sites} and \emph{bonds}, or \emph{actors} and \emph{ties}, respectively. Let $N$ be a network, and suppose that there is a map $w:L(N)\to \mathbb{R}$. Triple $(N(N),L(N),w)$ is called a \emph{weighted network}. 
  
A good introduction to the concept of networks can be found in \cite{Ne} and \cite{NBW}, whereas \cite{W} contains the same ideas described in the languange of graphs. Methods of weighted networks can be found in \cite{H}.  

There were $11$ jurors and $7$ contestants in the final stage of the $15^{\textrm{th}}$ Wieniawski Competition. The jurors rated the contestants in the final according to the Reverse Borda count: $1$ point for the best and $7$ points for the worst. The winner was the contenstant with the lower sum of points. The results are presented in Table \ref{tab:Wieniawski}.

\begin{figure}[ht]
\caption{Network $N(\textrm{\textbf{W}})$, green -- small weights, red -- large weights}\label{fig:Wieniawski}
\begin{center}
\includegraphics[width=0.7\textwidth]{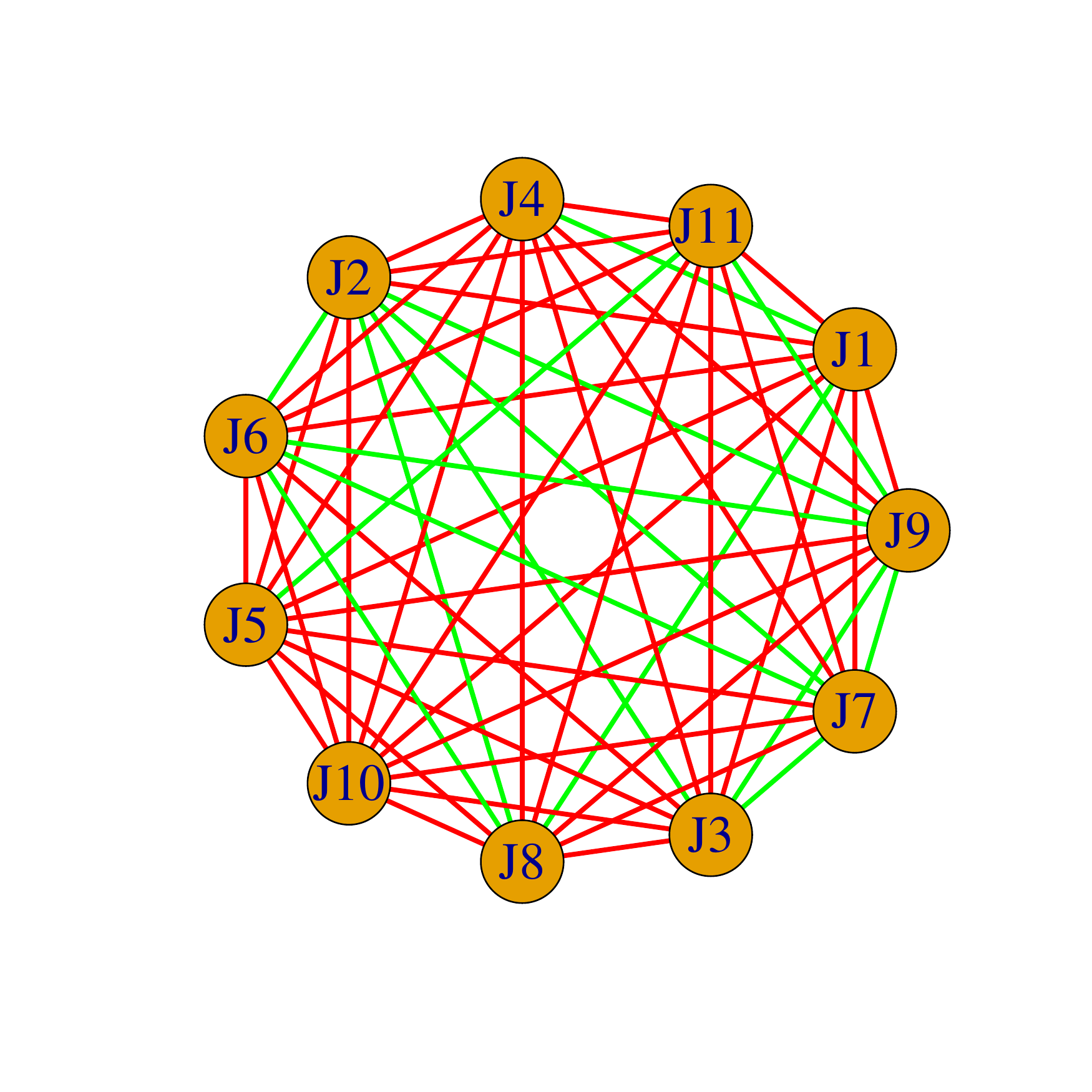}
\end{center}
\end{figure}

We define the weighted network $N(\textrm{\textbf{W}})$ in the following way. The nodes set of $N(\textrm{\textbf{W}})$ corresponds to the jurors set $\{J_{i}: i \in [11]\}$, whereas the links set consists of all links $\{J_{s},J_{t}:i\neq j \in [11]\}$. For link $l_{st}$ connecting $J_{s}$ with $J_{t}$, we assign weight $w(l_{st})=w_{st}$, where 
\begin{displaymath}
w_{st}=\mathcal{LF}_{2}(\alpha_{s}\alpha_{t}^{-1}) \textrm{.}
\end{displaymath}
Here $\alpha_{s}$ and $\alpha_{t}$ denote the votes of jurors $J_{s}$ and $J_{t}$, respectively. In particular, for $i=1,2,\ldots,7$, $\alpha_{s}(i)$ is the number of points given to the $i$-th contestant by juror $J_{s}$. The similar holds for juror $J_{t}$. Note that $\alpha_{s}$ and $\alpha_{t}$ can be considered as elements of $S_{7}$. Network $N(\textrm{\textbf{W}})$ is presented in Figure \ref{fig:Wieniawski}.

\begin{figure}[ht] 
\caption{Voting networks based on the final results of the $15^{\textrm{th}}$ International Henryk Wieniawski Violin Competition}\label{fig:WieniawskiModifiedI}
\begin{center}
\subfloat[Network $N(\textrm{\textbf{W}})_{0.4}$]{\includegraphics[width=7cm]{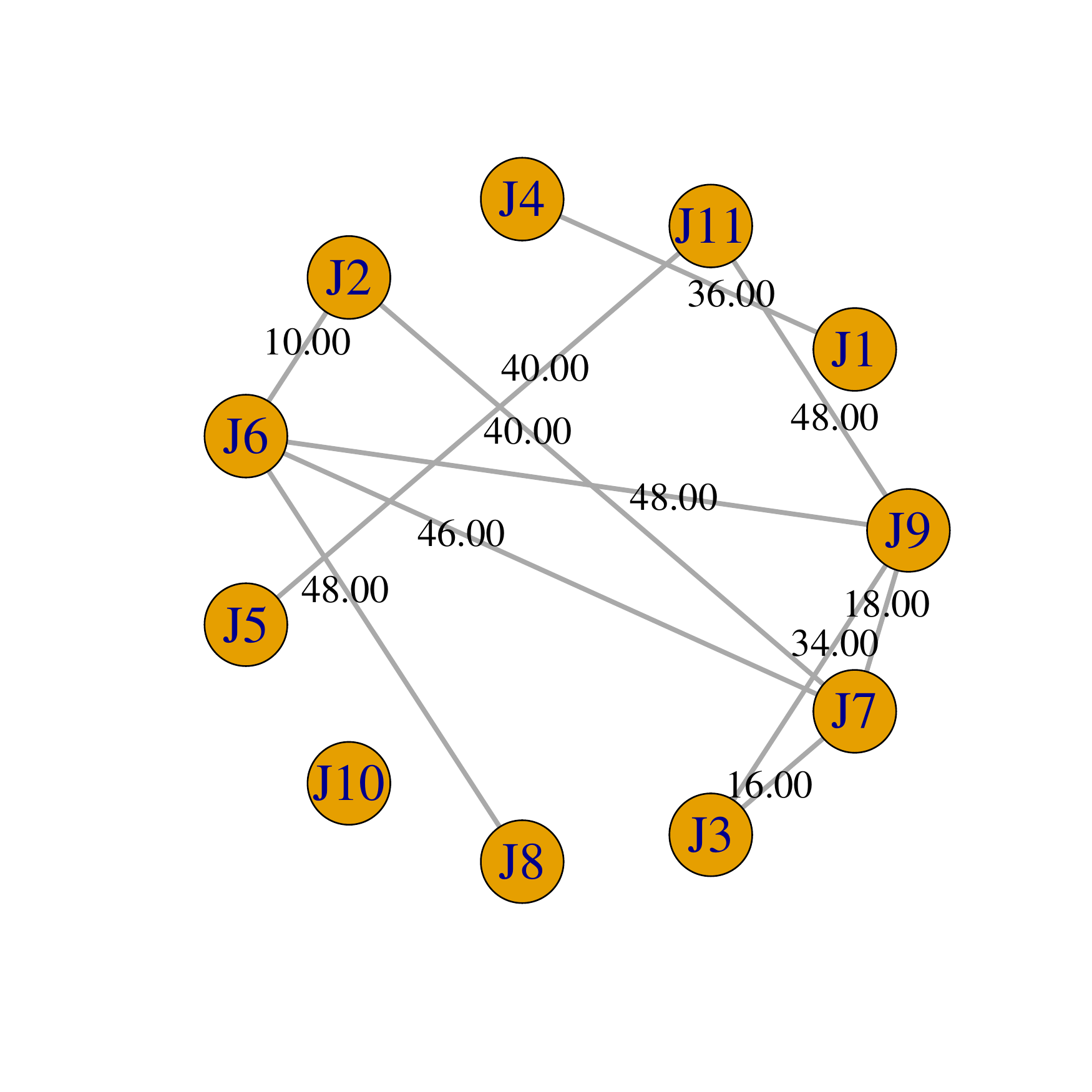}\label{fig:Wieniawski04}}
\quad
\subfloat[Network $N(\textrm{\textbf{W}})_{0.3}$]{\includegraphics[width=7cm]{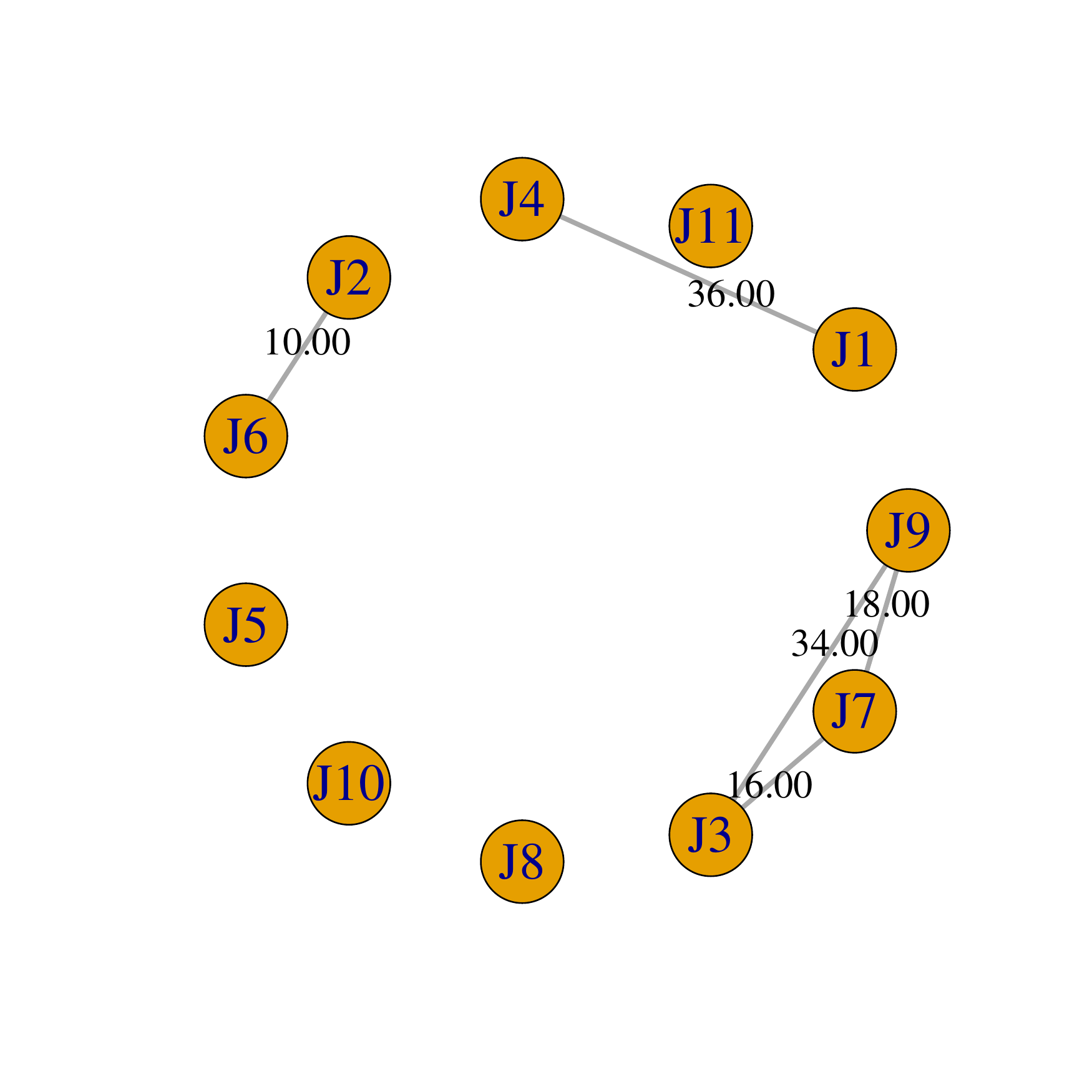}\label{fig:Wieniawski03}}
\\
\subfloat[Network $N(\textrm{\textbf{W}})_{0.2}$]{\includegraphics[width=7cm]{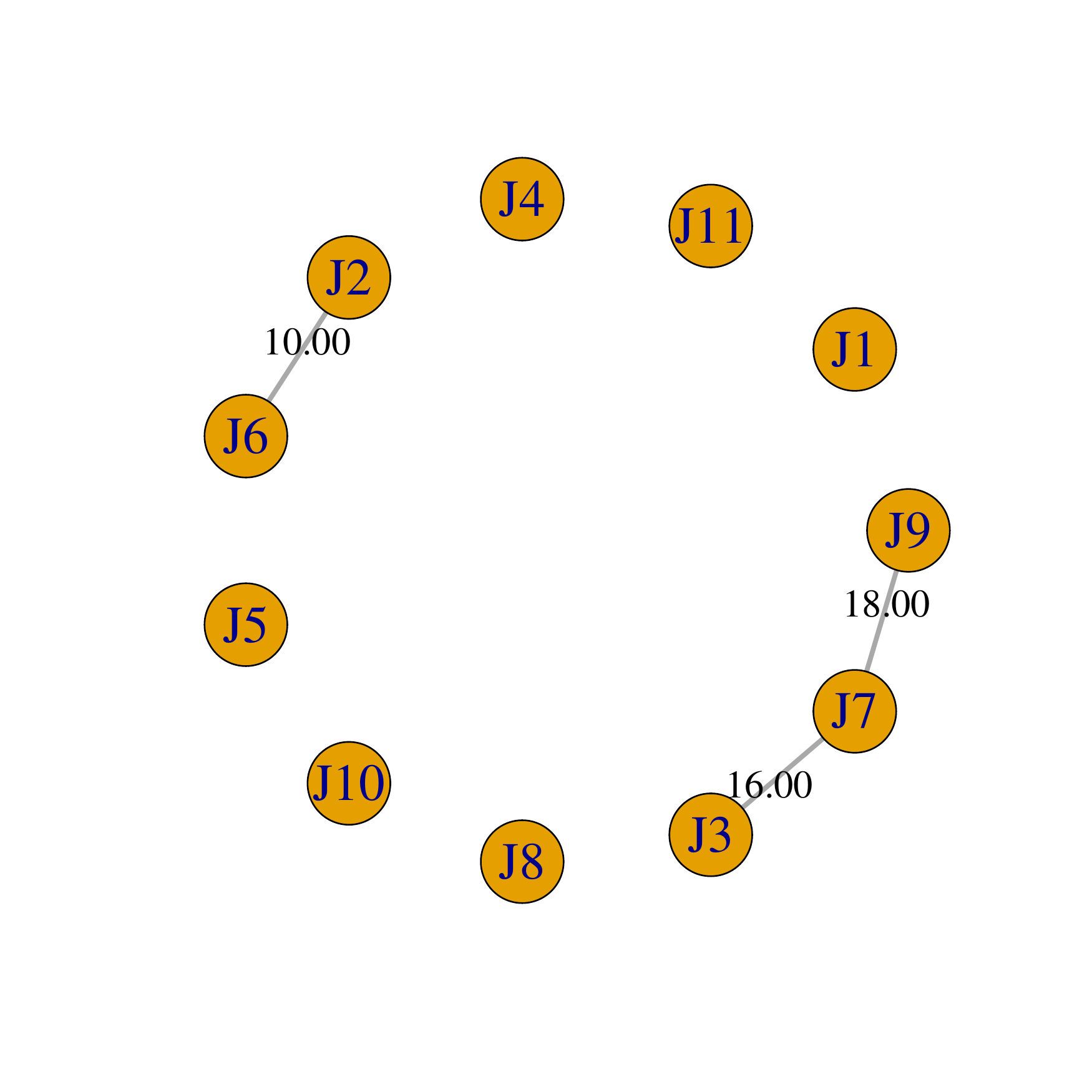}\label{fig:Wieniawski02}}
\quad
\subfloat[Network $N(\textrm{\textbf{W}})_{0.1}$]{\includegraphics[width=7cm]{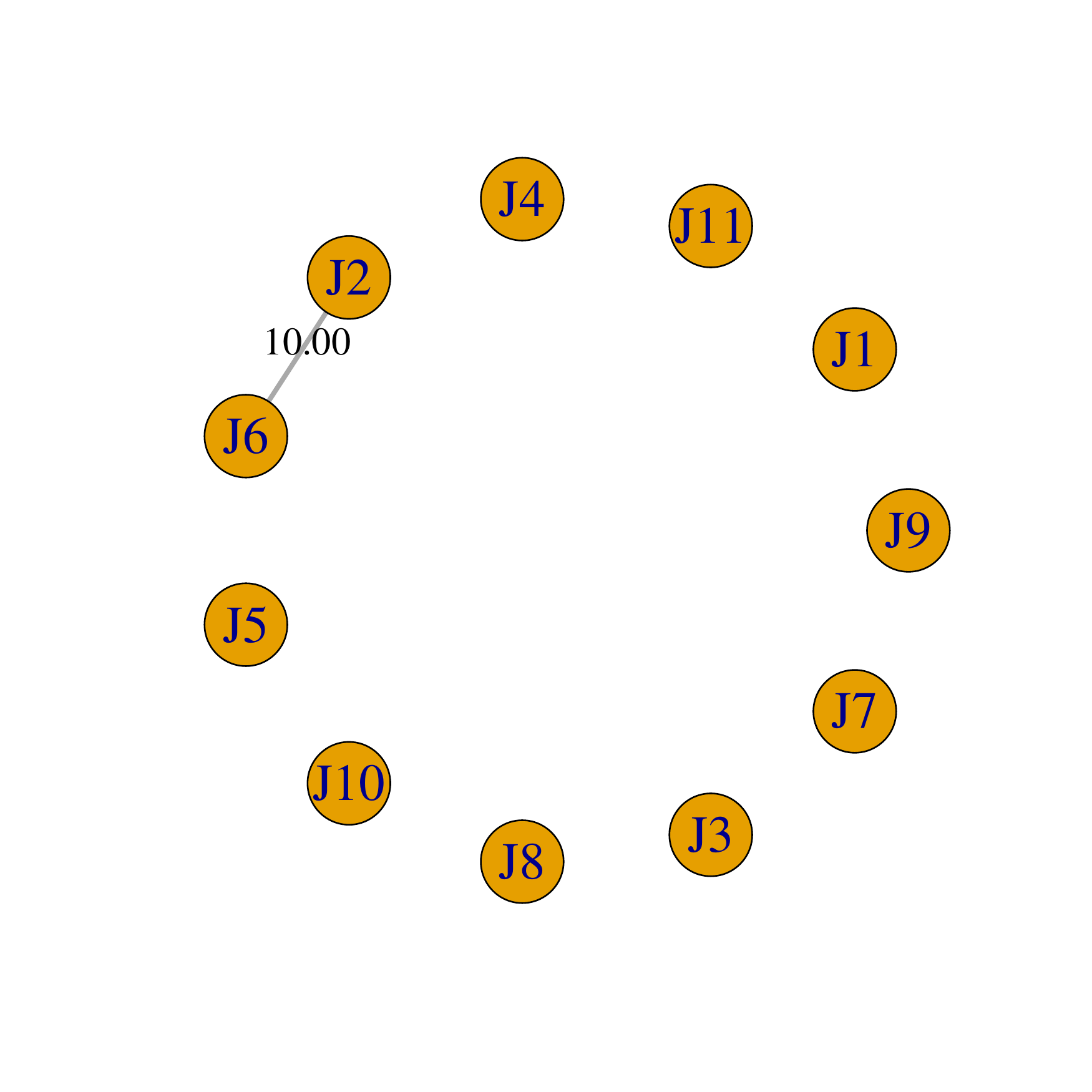}\label{fig:Wieniawski01}}
\end{center}
\end{figure}

For $p=1, 0.9, 0.8, 0.7, 0.6, 0.5, 0.4, 0.3, 0.2, 0.1, 0.0$, we create networks $N(\textrm{\textbf{W}})_{p}$ removing from $N(\textrm{\textbf{W}})$ links $l_{st}$ with weights satisfying the condition
\begin{displaymath}
w(l_{st}) > p \cdot \max{\{\mathcal{LF}_{2}(\sigma):\sigma \in S_{7}\}} = p \cdot 120 \textrm{.}
\end{displaymath} 
Note that for $p$ decreasing, $N(\textrm{\textbf{W}})_{p}$ contains links connecting jurors voting more and more consitently. They will be used later in Section \ref{sec:Fitting} . Networks $N(\textrm{\textbf{W}})_{0.4}$, $N(\textrm{\textbf{W}})_{0.3}$, $N(\textrm{\textbf{W}})_{0.2}$ and $N(\textrm{\textbf{W}})_{0.1}$ are presented in Figure \ref{fig:WieniawskiModifiedI}. 

\begin{figure}[ht] 
\caption{Voting networks based on the final results of the $15^{\textrm{th}}$ International Henryk Wieniawski Violin Competition}\label{fig:WieniawskiModifiedII}
\begin{center}
\subfloat[Network $N(\textrm{\textbf{W}})^{0.2}$]{\includegraphics[width=4cm]{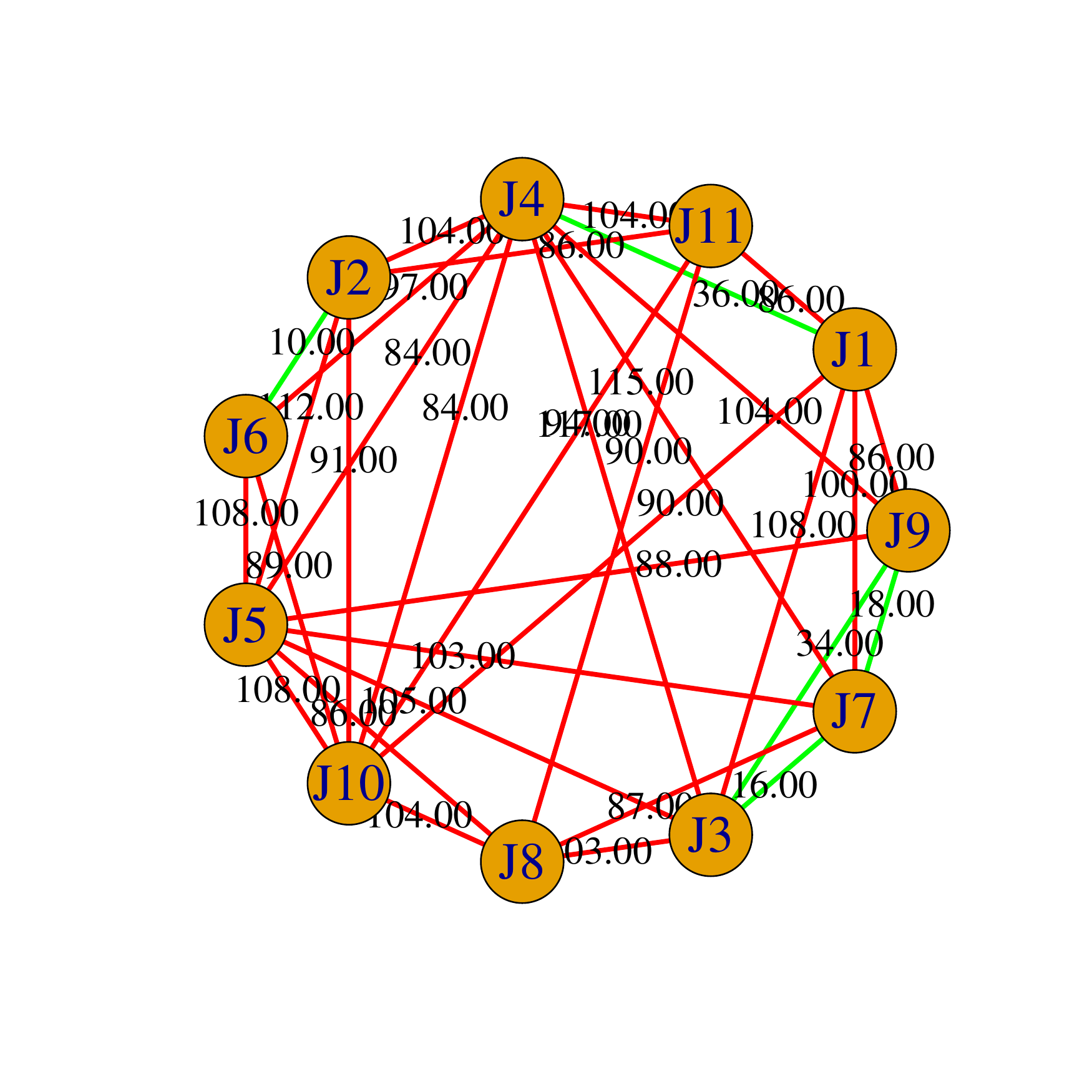}\label{fig:WieniawskiCol02}}
\quad
\subfloat[Network $N(\textrm{\textbf{W}})^{0.25}$]{\includegraphics[width=4cm]{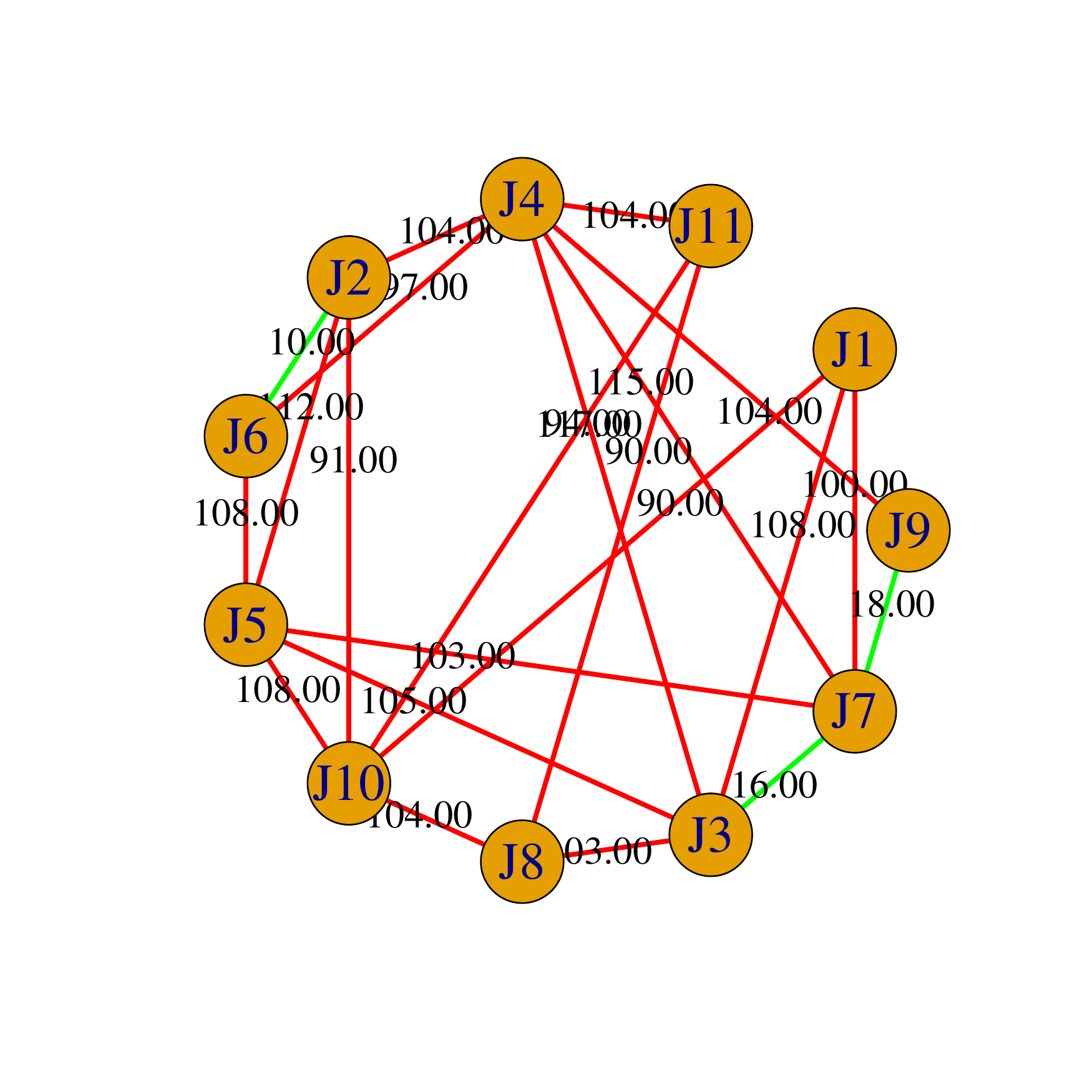}\label{fig:WieniawskiCol025}}
\quad
\subfloat[Network $N(\textrm{\textbf{W}})^{0.3}$]{\includegraphics[width=4cm]{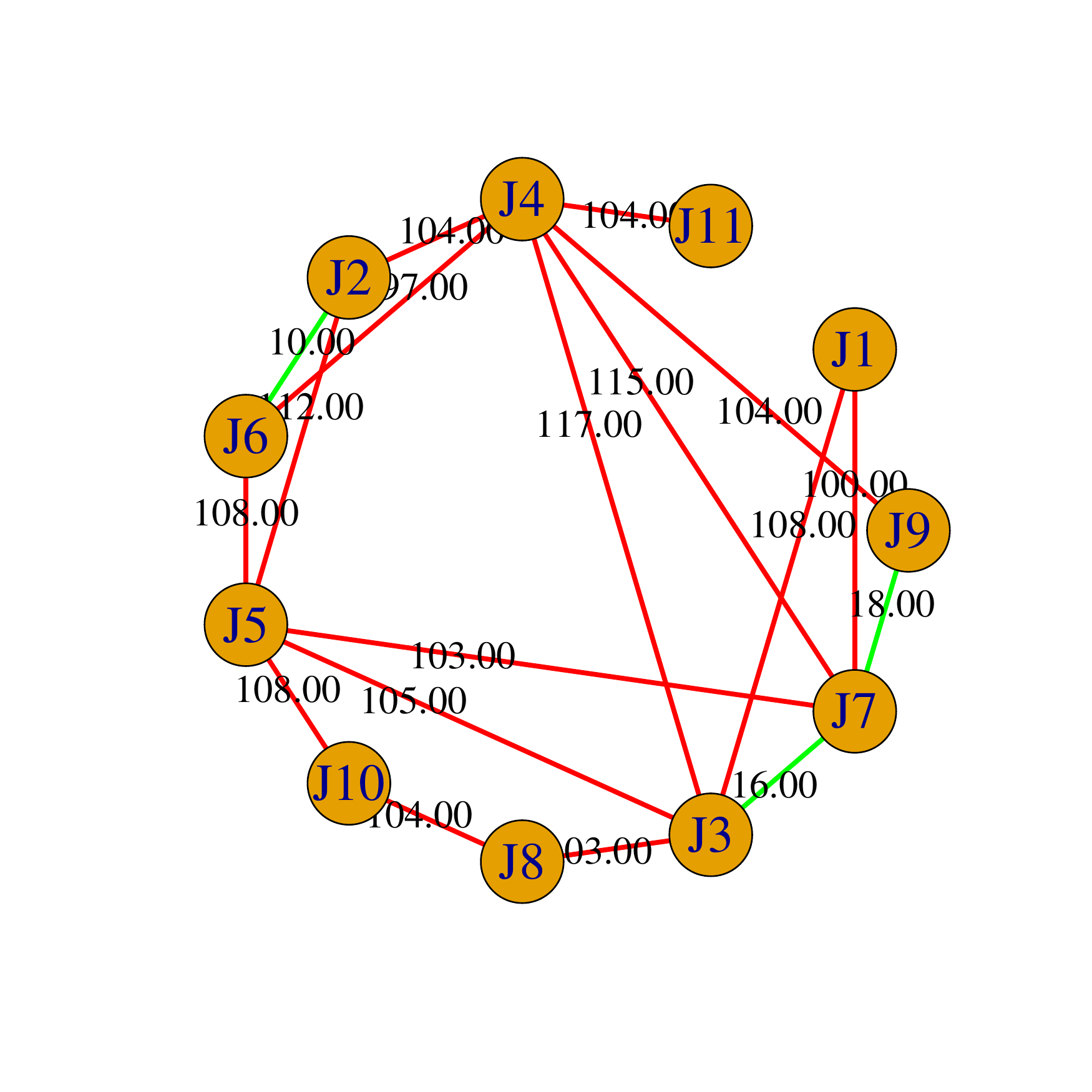}\label{fig:WieniawskiCol03}}
\\
\subfloat[Network $N(\textrm{\textbf{W}})^{0.35}$]{\includegraphics[width=4cm]{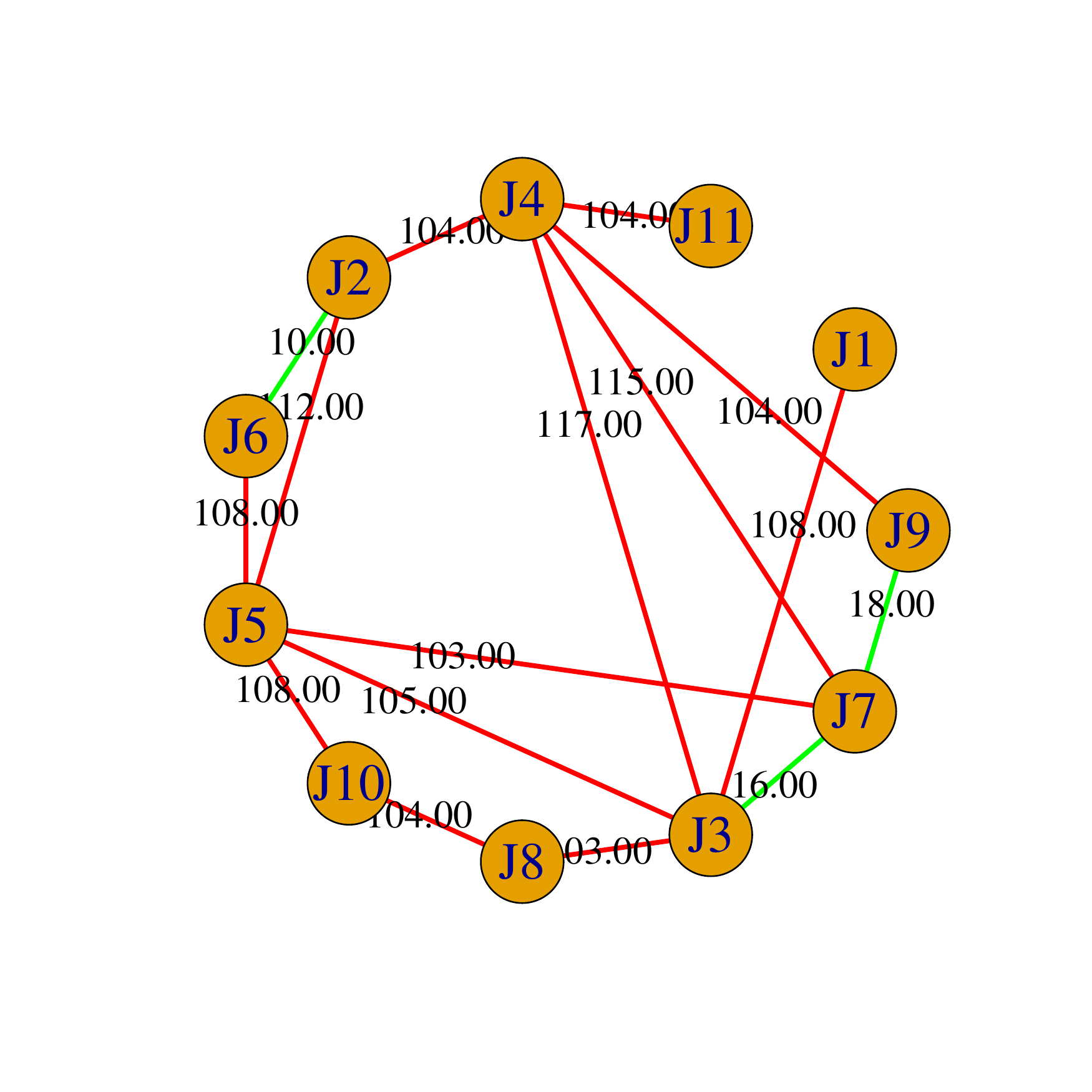}\label{fig:WieniawskiCol035}}
\quad
\subfloat[Network $N(\textrm{\textbf{W}})^{0.4}$]{\includegraphics[width=4cm]{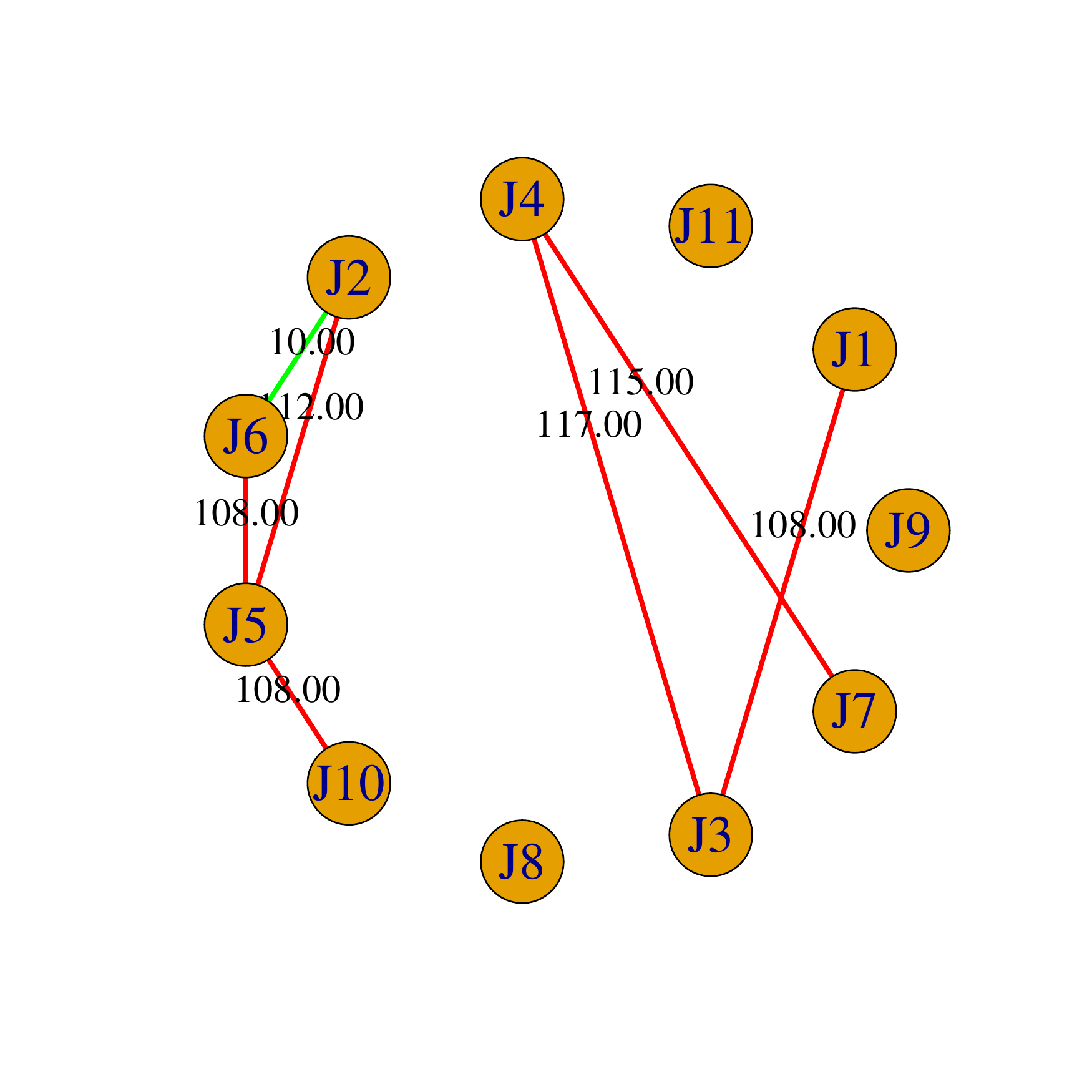}\label{fig:WieniawskiCol04}}
\quad
\subfloat[Network $N(\textrm{\textbf{W}})^{0.4}$]{\includegraphics[width=4cm]{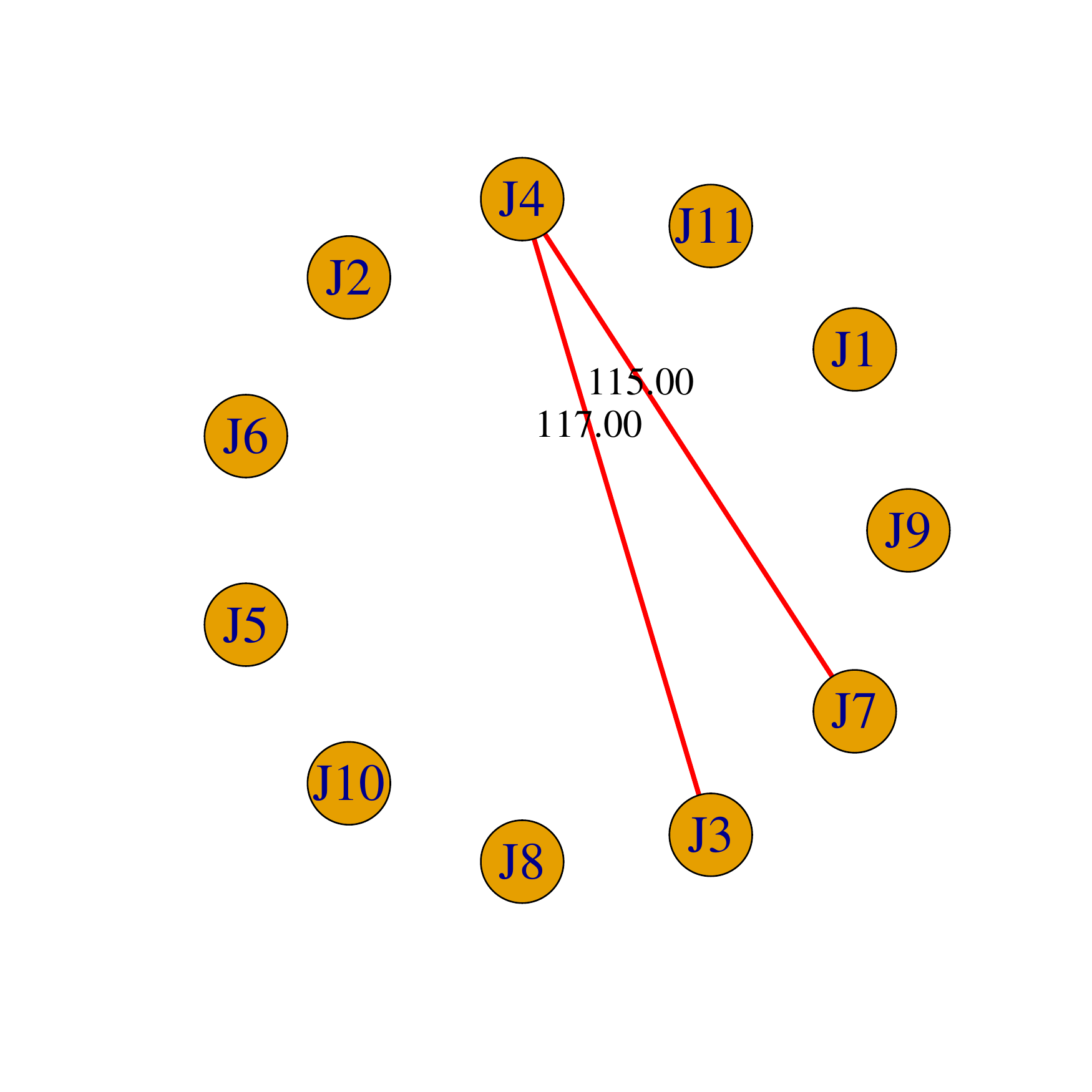}\label{fig:WieniawskiCol045}}
\end{center}
\end{figure}

For better understanding of the consistency of jurors' voting, for \\$s = 0,0.05,0.10,0.15,0.20,0.25,0.30,0.35,0.40,0.45,0.50$, we create networks $N(\textrm{\textbf{W}})^{s}$ removing from $N(\textrm{\textbf{W}})$ links $l_{st}$, with weights satisfying the condition
\begin{displaymath}
|w(l_{st}) - 0.5 \cdot \max{\{\mathcal{LF}_{2}(\sigma):\sigma \in S_{7}\}}| < s \cdot \max{\{\mathcal{LF}_{2}(\sigma):\sigma \in S_{7}\}} 
\textrm{, i.e.}
\end{displaymath}
\begin{displaymath}
|w(l_{st}) - 60| < s\cdot 120 \textrm{.}
\end{displaymath}
Note that when $s$ is increasing, then $N(\textrm{\textbf{W}})^{s}$ contains only the links connecting jurors that vote more and more consitently and more and more inconsistently, since we remove the links connecting jurors voting independently. Networks $N(\textrm{\textbf{W}})^{0.2}$, $N(\textrm{\textbf{W}})^{0.25}$, $N(\textrm{\textbf{W}})^{0.3}$, $N(\textrm{\textbf{W}})^{0.35}$, $N(\textrm{\textbf{W}})^{0.4}$ and $N(\textrm{\textbf{W}})^{0.45}$ are presented in Figure \ref{fig:WieniawskiModifiedII}.

\section{Generic networks of votings}\label{sec:Generic}

In this section, we describe the procedure of generating random networks of votings. These \emph{generic networks} will be used later in Section \ref{sec:Fitting} for determining the model of generating random networks with properties best fitting the properties of network $N(\textrm{\textbf{W}})$. We determine this best fitting model to check the hypothesis that jurors of $15^{\textrm{th}}$ International Henryk Wieniawski Violin Competition voted controversially. Namely, their votings were neither consistent nor random.

Consider the space of all possible votings of a single juror in \textbf{\textrm{W}} - the final stage of the $15^{\textrm{th}}$ Wieniawski Competition. This space can be seen as $S_{7}$. Let $\mathbb{P}$ denote the probability measure on $S_{7}$. We consider the measures of the form
\begin{displaymath}
\mathbb{P}=\mathbb{P}(d,\alpha,\beta)=\alpha\mathbb{P}_{\sigma_{1}}+\beta\mathbb{P}_{\sigma_{2}}+
(1-\alpha-\beta)\mathbb{P}_{\textrm{uniform}}
\end{displaymath}
where:
\begin{itemize}
\item $\alpha, \beta \in [0,1]$ satisfy the condition $\alpha + \beta \leq 1$,
\item $d \in [0,1]$,
\item $\mathbb{P}_{\textrm{uniform}}$ is the uniform probability measure on space $S_{7}$,
\item $\mathbb{P}_{\sigma_{1}}$ and $\mathbb{P}_{\sigma_{2}}$ are the Dirac probability measures on $S_{7}$ centred at permutations $\sigma_{1}$ and $\sigma_{2}$ respectively, where $\sigma_{1}$ and $\sigma_{2}$ are randomly chosen in such a way that they satisfy the condition
\begin{displaymath}
\mathcal{LF}_{2}(\sigma_{1}\sigma_{2}^{-1}) \simeq d\cdot \max{\{\mathcal{LF}_{2}(\sigma):\sigma \in S_{7}\}}\textrm{.}
\end{displaymath}
\end{itemize}

$\mathbb{P}$ defined in this way is a convex combination of $\mathbb{P}_{\sigma_{1}}$, $\mathbb{P}_{\sigma_{2}}$ and 
$\mathbb{P}_{\textrm{uniform}}$. The condition bonding $\sigma_{1}$ and $\sigma_{2}$ is a metric analogue of the condition for the correlation coefficient of $\sigma_{1}$ and $\sigma_{2}$. In this paper, we consider $\alpha,\beta = 0,0.05,0.1,\ldots,0.95,1$ and 
$d = 0,0.1,\ldots,0.9,1$. Set $N_{\textrm{max}}=\max{\{\mathcal{LF}_{2}(\sigma):\sigma \in S_{7}\}}$. The notation $\simeq$ means that $\sigma_{1}$ and $\sigma_{2}$ are randomly chosen from all possible pairs of permutations satisfying condition
$\mathcal{LF}_{2}(\sigma_{1}\sigma_{2}^{-1}) \in  \left[d\cdot N_{\textrm{max}} - 0.05\cdot N_{\textrm{max}}, d\cdot N_{\textrm{max}} + 0.05\cdot N_{\textrm{max}} \right]$.

The procedure for generating the random network of votings is as follows:
\begin{enumerate}[(i)]
\item choose a repetition number $j=1,2,\ldots,100$,
\item choose $d = 0,0.1,\ldots,0.9,1$,
\item randomly choose such $\sigma_{1}$ and $\sigma_{2}$ that $\mathcal{LF}_{2}(\sigma_{1}\sigma_{2}^{-1}) \simeq d\cdot N_{\textrm{max}}$,
\item choose such $\alpha,\beta = 0,0.05,0.1,\ldots,0.95,1$ that $\alpha+\beta \leq 1$,
\item for every $s = 1,2,\ldots,11$ randomly, according to $\mathbb{P}=\alpha\mathbb{P}_{\sigma_{1}}+\beta\mathbb{P}_{\sigma_{2}}+
(1-\alpha-\beta)\mathbb{P}_{\textrm{uniform}}$, choose $\alpha_{n}^{s}\in S_{7}$ -- this is the vote of juror $J_{s}$,
\item create weighted network $N(d,\alpha,\beta,j)$ according to the procedure described in Section \ref{sec:Wieniawski},
\item for $p=1, 0.9, 0.8, 0.7, 0.6, 0.5, 0.4, 0.3, 0.2, 0.1, 0.0$, create network $N(d,\alpha,\beta,j)_{p}$.
\end{enumerate}

\section{Fitting the parameters}\label{sec:Fitting}
In this section, we determine such parameters $d$, $\alpha$ and $\beta$ that the statistical properties of a family of networks 
$\{N(d,\alpha,\beta,j):j=1,2,\ldots,100\}$ fit best the properties of the network $N(\textrm{\textbf{W}})$. 

According to the constructions of networks $N(\textrm{\textbf{W}})_{p}$, their connected components, as $p$ decreases, correspond to groups of jurors voting in a more and more consistent way. Table \ref{tab:ConnectedComponents} contains the number of connected components of networks $N(\textrm{\textbf{W}})_{p}$.

Let $\textrm{C}(N)$ denote the number of connected components of network $N$. We determine the numbers $d_{\textrm{e}}^{\textrm{min}}$, 
$d_{\textrm{m}}^{\textrm{min}}$, $d_{\textrm{e}}^{\textrm{max}}$ and $d_{\textrm{m}}^{\textrm{max}}$ given by
\begin{displaymath}
d_{\textrm{e}}^{\textrm{min}} = \min_{d,\alpha,\beta}\left\{\sqrt{\sum_{p}\left(\textrm{C}\left(N(\textrm{\textbf{W}})_{p}\right) - 
\frac{\sum_{j = 1}^{100}\textrm{C}(N(d,\alpha,\beta,j)_{p})}{100}\right)^{2}}\right\} \textrm{,}
\end{displaymath}
\begin{displaymath}
d_{\textrm{m}}^{\textrm{min}} = \min_{d,\alpha,\beta}\left\{\sum_{p}\left|\textrm{C}\left(N(\textrm{\textbf{W}})_{p}\right) - 
\frac{\sum_{j = 1}^{100}\textrm{C}(N(d,\alpha,\beta,j)_{p})}{100}\right|\right\}\textrm{,}
\end{displaymath}
\begin{displaymath}
d_{\textrm{e}}^{\textrm{max}} = \max_{d,\alpha,\beta}\left\{\sqrt{\sum_{p}\left(\textrm{C}\left(N(\textrm{\textbf{W}})_{p}\right) - 
\frac{\sum_{j = 1}^{100}\textrm{C}(N(d,\alpha,\beta,j)_{p})}{100}\right)^{2}}\right\} \textrm{ and }
\end{displaymath}
\begin{displaymath}
d_{\textrm{m}}^{\textrm{max}} = \max_{d,\alpha,\beta}\left\{\sum_{p}\left|\textrm{C}\left(N(\textrm{\textbf{W}})_{p}\right) - 
\frac{\sum_{j = 1}^{100}\textrm{C}(N(d,\alpha,\beta,j)_{p})}{100}\right|\right\}
\end{displaymath}
respectively.

Note that $d_{\textrm{e}}^{\textrm{min}}$ and $d_{\textrm{m}}^{\textrm{min}}$ minimalise the Euclidean distance and the Manhattan distance between the number of connected components of $N(\textrm{\textbf{W}})_{p}$ and the average number of connected components of 
$N(d,\alpha,\beta,j)_{p}$, respectively. Similarily, $d_{\textrm{e}}^{\textrm{max}}$ and $d_{\textrm{m}}^{\textrm{max}}$ maximalise
these distances. Table \ref{tab:Parameters} contains parameters $d$, $\alpha$ and $\beta$ of models for which these distances are the smallest. The sum
\begin{displaymath}
\sqrt{\sum_{p}\left(\textrm{C}\left(N(\textrm{\textbf{W}})_{p}\right) - 
\frac{\sum_{j = 1}^{100}\textrm{C}(N(d,\alpha,\beta,j)_{p})}{100}\right)^{2}}
\end{displaymath} 
varies between $d_{\textrm{e}}^{\textrm{min}}=14.29699$ and $d_{\textrm{e}}^{\textrm{max}}=24.75912$, whereas the sum
\begin{displaymath}
\sum_{p}\left|\textrm{C}\left(N(\textrm{\textbf{W}})_{p}\right) - 
\frac{\sum_{j = 1}^{100}\textrm{C}(N(d,\alpha,\beta,j)_{p})}{100}\right|
\end{displaymath}
varies between $d_{\textrm{m}}^{\textrm{min}}=34.28$ and $d_{\textrm{m}}^{\textrm{max}}=75.02$. In both cases best fitted models are those with parameter $d=1$. Note that $d=1$ means that $\sigma_{1}$ and $\sigma_{2}$, chosen during the procedure described in Section \ref{sec:Generic}, satisfy $\mathcal{LF}_{2}(\sigma_{1}\sigma_{2}^{-1}) \simeq N_{max}$, therefore being almost revers permutations. 
This fact confirms the hypothesis that jurors of the $15^{\textrm{th}}$ International Henryk Wieniawski Violin Competition were far away from being consistent.

\section{Conclusions and recommendations for further research}\label{sec:Conclusions}
This section contains conclusions and open questions related to the subject of these studies.

The results of voting include a lot of information about preferences of voters and their structure. The application of network theory can highlight properties of networks constructed on the basis of jurors' votings. The obtained networks may be used to describe homogeneity or heterogeneity of jurors' votings. 

During this research some questions arose. 
\begin{enumerate}
\item How do statistical coefficients of networks $N(d,\alpha,\beta,j)_{p}$ depend on $d$, $\alpha$ and $\beta$?
\item How does behaviour of generic networks depend on the number of jurors and the number of contestants?
\item What are the asymptotic (with number of jurors and/or contestants tending to $\infty$) properties of generic networks?
\item How do best fitting parameters $d$, $\alpha$ and $\beta$ change when increasing the number of repeats? 
\item How will properties of generic networks change when we randomly choose permutation $\sigma_{1}$ and $\sigma_{2}$ with 
$\mathcal{LF}_{2}(\sigma_{1}\sigma_{2}^{-1})$ precisely set from all possible values of $\mathcal{LF}_{2}$?
\end{enumerate}
These questions are a good starting point for further research.

\subsection*{Acknowledgements}

While working on this paper, the second author was partially supported by the SGH fund KAE/S21 and by the NCN fund UMO-2018/31/B/HS4/01005.

All calculations and figures were prepared with R 4.0.3, see\cite{R}.

The second author is grateful to his whife for her strong and loving support as well as for her inspiring \emph{''try to think nonstandard''}.

\section{Tables and Figures}\label{sec:Tables}

\begin{table}[ht]
\caption{Final results of the $15^{\textrm{th}}$ International Henryk Wieniawski Violin Competition}\label{tab:Wieniawski}
\begin{center}
\begin{tabular}{|c|c|c|c|c|c|c|c|c|c|c|c|}
\hline
  & J1 & J2 & J3 & J4 & J5 & J6 & J7 & J8 & J9 & J10 & J11 \\
\hline
A & $7$ & $3$ & $2$ & $7$ & $7$ & $4$ & $3$ & $7$ & $7$ & $7$ & $7$ \\
\hline
B & $4$ & $7$ & $7$ & $2$ & $2$ & $7$ & $7$ & $2$ & $5$ & $6$ & $5$ \\
\hline
C & $5$	& $5$ &	$5$ & $3$ & $6$ & $6$ &	$5$ & $5$ & $6$ & $1$ & $6$ \\
\hline
D & $3$	& $6$ &	$4$ & $5$ & $1$ & $5$ & $4$ & $4$ & $3$ & $5$ &	$1$ \\
\hline
E & $1$ & $4$ &	$6$ & $1$ & $3$	& $3$ & $6$ & $3$ & $4$ & $3$ &	$4$ \\
\hline
F & $6$ & $2$ & $1$ & $6$ & $4$ & $2$ &	$1$ & $6$ & $1$	& $2$ & $2$ \\
\hline
G & $2$ & $1$ & $3$ & $4$ & $5$ & $1$ & $2$ & $1$ & $2$ & $4$ & $3$ \\ 
\hline
\end{tabular}
\end{center}
\end{table}

\begin{table}[ht]
\caption{Number of connected componets of networks $N(\textrm{\textbf{W}})_{p}$}\label{tab:ConnectedComponents}
\begin{center}
\begin{tabular}{|c|c|}
\hline
treshold parameter $p$ & component number \\
\hline
1 & 1 \\
\hline
0.9 & 1 \\
\hline
0.8 & 1 \\
\hline
0.7 & 1 \\
\hline
0.6 & 2 \\
\hline
0.5 & 2 \\
\hline
0.4 & 3 \\
\hline
0.3 & 7 \\
\hline
0.2 & 8 \\
\hline
0.1 & 10 \\
\hline
0 & 11 \\
\hline
\end{tabular}
\end{center}
\end{table}

\begin{table}[ht]
\caption{Best fitting parameters}\label{tab:Parameters}
\begin{center}
\begin{tabular}{|c|c|}
\hline
Euclidean Distance & maximal value - $24.75912$ \\
\hline
$d=1$, $\alpha=0.4$, $\beta=0.6$ & $14.29699$\\
\hline
$d=1$, $\alpha=0.45$, $\beta=0.55$ & $14.29773$\\
\hline
$d=1$, $\alpha=0.5$, $\beta=0.5$ & $14.30108$\\
\hline
$d=1$, $\alpha=0.55$, $\beta=0.45$ & $14.30696$ \\
\hline
$d=1$, $\alpha=0.6$, $\beta=0.4$ & $14.30696$\\
\hline
Manhattan Distance & maximal value - $75.02$ \\
\hline
$d=1$, $\alpha=0.1$, $\beta=0.9$ & $34.28$ \\
\hline
$d=1$, $\alpha=0.15$, $\beta=0.85$ & $34.32$ \\
\hline
$d=1$, $\alpha=0.85$, $\beta=0.15$ & $34.32$\\
\hline
$d=1$, $\alpha=0.9$, $\beta=0.1$ & $34.4$ \\
\hline
$d=0.9$, $\alpha=0.1$, $\beta=0.9$ & $34.78$ \\
\hline
\end{tabular}
\end{center}
\end{table}

\end{document}